\documentclass[a4paper,fleqn]{cas-dc}

\usepackage[numbers]{natbib}

\def\tsc#1{\csdef{#1}{\textsc{\lowercase{#1}}\xspace}}
\tsc{WGM}
\tsc{QE}
\tsc{EP}
\tsc{PMS}
\tsc{BEC}
\tsc{DE}

\begin{document}
\let\WriteBookmarks\relax
\def\floatpagepagefraction{1}
\def\textpagefraction{.001}

\shorttitle{}

\shortauthors{Tian et~al.}

\title [mode = title]{Multi-Depth Branch Network for Efficient Image Super-Resolution}                      
\tnotemark[1]

\tnotetext[1]{This work is the results of the research
   project funded by the National Science Foundation of China under Grants 61925603. }

%
\author[1]{Huiyuan Tian}


\ead{tianhuiyuan@zju.edu.cn}


\affiliation[1]{organization={College of Computer Science and Technology, Zhejiang University},
    addressline={NO. 38 Zheda Road, Xihu District}, 
    city={Hangzhou},
    postcode={310027}, 
    country={China}}

\affiliation[2]{organization={Advanced Technology Research Institute, Zhejiang University},
    addressline={NO. 38 Zheda Road, Xihu District}, 
    city={Hangzhou},
    postcode={310027}, 
    country={China}}

\author[1,2]{Li Zhang}
\ead{zhangli85@zju.edu.cn}
\author[1]{Shijian Li}
\cormark[1]
\ead{shijianli@zju.edu.cn}

\author[1]{Min Yao}
\ead{myao@zju.edu.cn}
\author[1]{Gang Pan}
\ead{gpan@zju.edu.cn}

\cortext[cor1]{Corresponding author at: College of Computer Science and Technology, Zhejiang University, NO. 38 Zheda Road, Xihu District, Hangzhou,  310027, Zhejiang, China}



\begin{abstract}
A longstanding challenge in Super-Resolution (SR) is how to efficiently enhance high-frequency details in Low-Resolution (LR) images while maintaining semantic coherence. This is particularly crucial in practical applications where SR models are often deployed on low-power devices. To address this issue, we propose an innovative asymmetric SR architecture featuring Multi-Depth Branch Module (MDBM). These MDBMs contain branches of different depths, designed to capture high- and low-frequency information simultaneously and efficiently. The hierarchical structure of MDBM allows the deeper branch to gradually accumulate fine-grained local details under the contextual guidance of the shallower branch.  
We visualize this process using feature maps, and further demonstrate the rationality and effectiveness of this design using proposed novel Fourier spectral analysis methods. Moreover, our model exhibits more significant spectral differentiation between branches than existing branch networks. This suggests that MDBM reduces feature redundancy and offers a more effective method for integrating high- and low-frequency information. Extensive qualitative and quantitative evaluations on various datasets show that our model can generate structurally consistent and visually realistic HR images. It achieves state-of-the-art (SOTA) results at a very fast inference speed. Our code is available at \url{https://github.com/thy960112/MDBN}.

\end{abstract}



\maketitle

\section{Introduction}

Single Image Super-Resolution (SISR) is a crucial task in the field of image processing, aiming to reconstruct High-Resolution (HR) images from their LR counterparts \cite{wang2020deep, moser2023hitchhiker}. With the rapid development of deep learning, significant progress has been made in the field of SR.

As a pixel-wise prediction task, how to efficiently enhance high-frequency details in LR images while maintaining semantic coherence has been a longstanding challenge in SR. A variety of deep learning approaches have been applied to address this challenge. Dong et al. \cite{dong2015SRCNN} pioneered the use of CNNs for SR tasks. Subsequently, tremendous research \cite{dong2016accelerating, zhang2018image} began to focus on how to use CNNs to achieve better SR results, including the use of generative models such as Generative Adversarial Networks (GANs) \cite{ledig2017SRGAN, wang2018ESRGAN, wang2021RealESRGAN, wu2023IMAVISGAN} and diffusion models \cite{li2022diffsr, saharia2022SR3, gao2023ddimsr}. Several SR methods \cite{kim2016VDSR, 2017EDSRBSFlickr2K, zhang2018residual, kong2022residual, li2022blueprint} use extensive layer stacks and utilize residual connections \cite{he2016ResNet} to stabilize the training process.

In recent developments, some methods \cite{liang2021swinir, lu2022transformersr, zhang2022ELAN,  liu2023coarse, zhou2023srformer} have incorporated the transformer architecture \cite{2017Transformer, dosovitskiy2020ViT, liu2021swin, xu2023ssl} into SR tasks. This approach has garnered considerable attention in the research community, primarily due to its ability to capture long-range dependencies within images. However, this capability often leads to a substantial increase in computational cost and the number of parameters. As a result, there is a growing trend towards the development of more efficient SR methods. Nevertheless, it is noteworthy that, compared to their CNN-based counterparts, the inference speed of these lightweight SR transformers \cite{liang2021swinir, zhang2022ELAN,  liu2023coarse} usually remains significantly lower by an order of magnitude.

\begin{figure}[!t]
	\centering
	\includegraphics[width=0.49\textwidth]{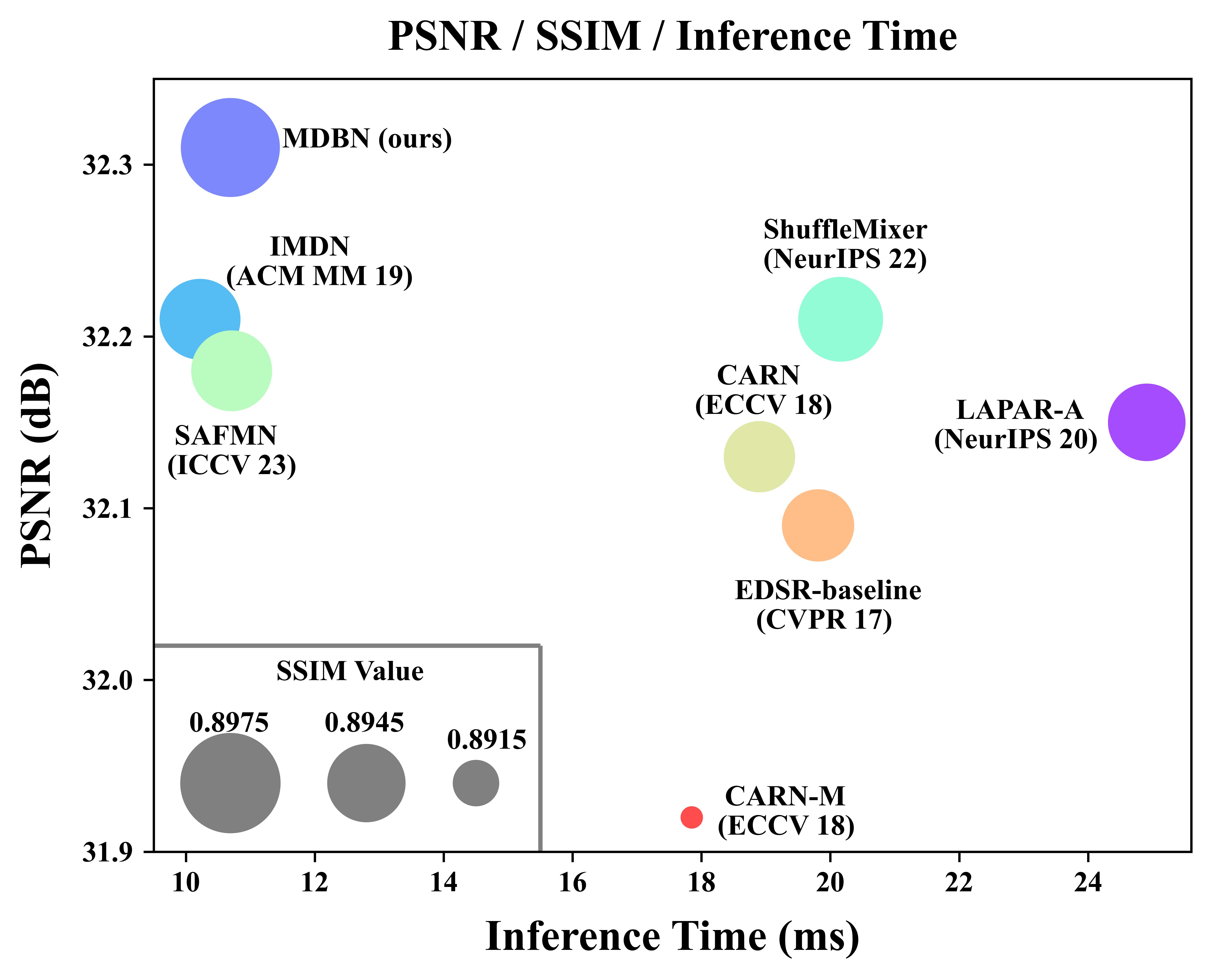} 
	\caption{Comparisons of PSNR, SSIM and inference time between our proposed MDBN model and other efficient methods for $\times4$ SR. The size of circles in the graph represents the SSIM value. Both PSNR and SSIM are evaluated on Set5 benchmark \cite{2012Set5}. Inference time is the average test time over a set of 50 LR images with dimensions of $320 \times 180$ pixels, using an NVIDIA RTX 3090 GPU.}
	\label{MDBNPSNRSSIM}
\end{figure}

In practical applications, SR models are frequently deployed on low-power devices, such as smartphones, which impose high demands on the inference speed of the SR models. To achieve efficient SR, we propose an innovative asymmetric architecture consisting of Multi-Depth Branch Modules (MDBMs). This module is composed of branches with different depths, all equipped with the $3\times3$ convolutional kernels, which is more computationally efficient. The motivation behind this asymmetric design is to leverage the deeper branch for high-frequency information restoration and fine-grained details capture, while the shallower branch concentrates on low-frequency information and outlines broader object contours.

To validate the logic and efficacy of our design, we employ feature map visualizations along with Fourier spectral analysis. These analyses reveal that the asymmetric branch network adeptly distinguishes lower-frequency structural elements from higher-frequency textural details while proficiently merging these information. This hierarchical configuration allows one branch to gradually gather fine-grained local details under the guidance of the wider contextual understanding contributed by the other branch, thereby demonstrating a synergistic approach to detail enhancement and coherence in SR tasks.

Our model exhibits more significant spectral differentiation between branches than existing branch networks that use identical-depth branches but varying convolutional kernel sizes. This observation highlights the capability of the MDBM to diminish feature redundancy and to serve as a more effective mechanism for the amalgamation of high- and low-frequency information.

This network design, which is specifically tailored for SR tasks, enables the model to generate structurally coherent and visually realistic HR images efficiently. As illustrated in Fig.~\ref{MDBNPSNRSSIM}, our proposed framework achieves SOTA results at a comparable inference speed compared to efficient SR networks. 

Our contributions in this paper can be summarized as follows:
\begin{itemize}
\item We propose a Multi-Depth Branch Network (MDBN), a simple yet powerful CNN-based network for efficient SR, which achieves comparable performance and significantly faster inference speed, ranging from $1.2\times$ to $18\times$, compared to lightweight SR transformers. It outperforms other SOTA efficient SR methods.

\item We design a novel asymmetric MDBM to capture and fuse high- and low-frequency information simultaneously and efficiently. The hierarchical design of MDBM allows the deeper branch to gradually accumulate detailed local features, guided contextually by the shallower branch.

\item We propose a Fourier spectral analysis method for feature maps to prove rationality and effectiveness of the MDBM. The more significant difference in the Fourier spectra between branches in MDBM indicates its ability to minimize feature redundancy and improve the overall efficiency of feature extraction.
\end{itemize}

\begin{figure*}[!t]
	
	\centering
	\includegraphics[width=0.99\textwidth]{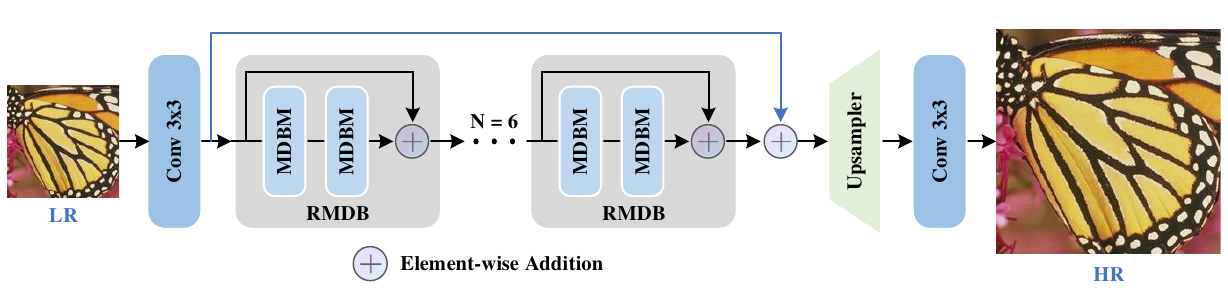} 
	\caption{An overview of the proposed MDBN. The MDBN first converts the LR image inputs into feature space utilizing an initial convolutional layer. Next, it employs a set of Residual Multi-Depth Branches (RMDBs) blocks to extract features, followed by an upsampler module for image reconstruction. The RMDB block consists of two MDBMs layers.}
	\label{MDBNarc}
\end{figure*}

\begin{figure*}[!t]
	\centering
	\includegraphics[width=0.99\textwidth]{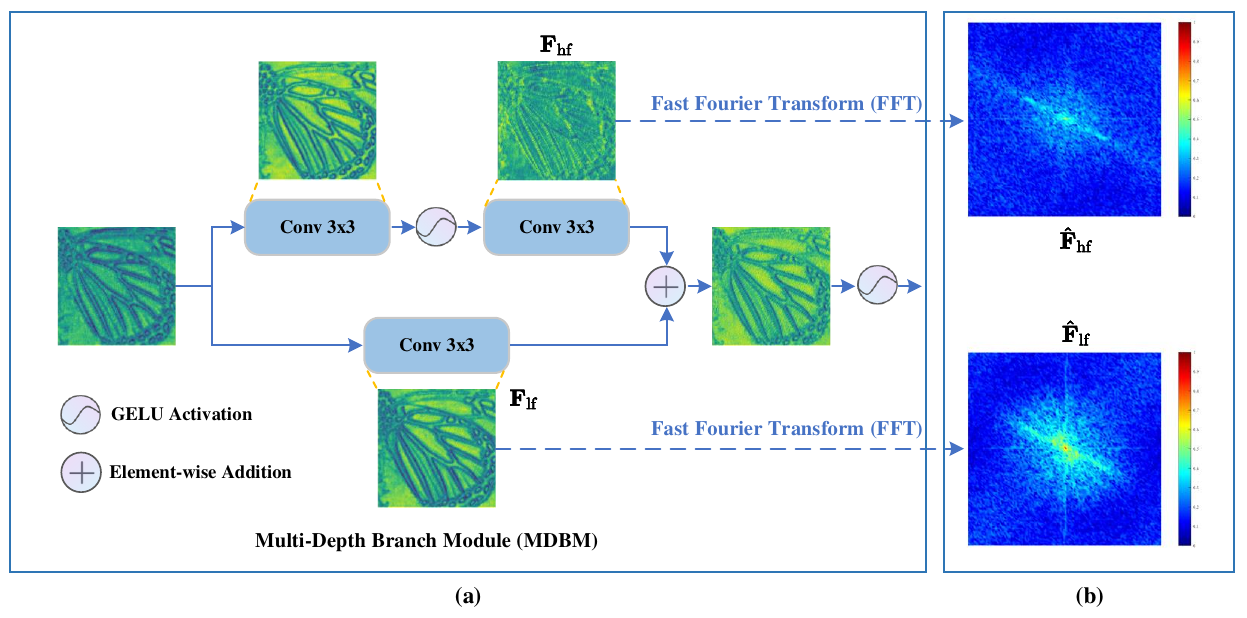} 
	\caption{(a) The details of the proposed Multi-Depth Branch Module (MDBM) and its corresponding feature map visualizations on $\times2$ SR. (b) Normalized Fourier spectra of feature maps from different branches. The region with larger values in the center shows more low-frequency content. It can be observed from $\hat{\mathbf{F}}_{\mathrm{hf}}$ that the deeper branch mainly predicts the high-frequency information and captures fine-grained object details. Meanwhile, $\hat{\mathbf{F}}_{\mathrm{lf}}$ reveals that the shallower branch contains more low-frequency semantic understanding and delineates broader object contours.}
	\label{MDBNFM}
\end{figure*}

\section{Related work}

\subsection{Networks for efficient SR}

Since the introduction of CNNs to the field of image SR by SRCNN \cite{dong2015SRCNN}, which marked a significant performance improvement over traditional SR methods, tremendous methods \cite{wang2023IMAVISlightweight, gao2023ESWA, chen2023hat, zhou2023srformer, shamsolmoali2019IMAVISDENSE, liu2023coarse, chen2021pre, kong2022residual, li2022blueprint, michelini2022edge} have been developed to improve the quality of image reconstruction. In real-time applications, there is a critical requirement for SR models to possess rapid inference speeds. This necessity has spurred research efforts towards developing efficient SR networks.

FSRCNN \cite{dong2016accelerating} and ESPCN \cite{Shi2016ESPCN} employ a post-upscaling approach, significantly reducing computational load by handling pre-defined inputs more efficiently. The LapSRN \cite{lai2017LapSRN} approach combines deep learning with the concept of the Laplacian pyramid to predict high-resolution details at different scales. EDSR-baseline \cite{2017EDSRBSFlickr2K} is designed to enhance image super-resolution using a deep residual network. Interestingly, it challenges the necessity of incorporating BatchNorm (BN) layers for SR tasks, indicating that such layers are not indispensable in this context. CARN \cite{ahn2018CARN} adopts a cascading structure of residual blocks to effectively capture intricate image details. IMDN \cite{hui2019IMDN} introduces a novel approach to super-resolution that utilizes iterative processing and multiple domains to enhance image details. ShuffleMixer \cite{sun2022ShuffleMixerNIPS} introduces large kernel convolutions for efficient SR. SAFMN \cite{sun2023SAFM} presents a spatially-adaptive feature modulation proved to be effective for enhancing image super-resolution. Despite these advancements addressing various aspects of efficiency, there remains a significant scope for optimizing the balance between reconstruction performance and model efficiency in SR applications.

\subsection{Multi-branch networks}

Multi-branch architectures \cite{szegedy2015goo, han2022multi, tian2023pyramid} are frequently employed to effectively capture diverse features. Such multi-branch architecture is found to be less non-convex in terms of duality gap~\cite{zhang2019deep}, indicating that they are more straightforward to train. A notable early example is GoogLeNet \cite{szegedy2015goo}, which incorporates parallel branches to process features at various scales and resolutions. DenseNet~\cite{huang2017densely} represents another form of multi-branch design. In this approach, each layer is interconnected with all subsequent layers, promoting feature reuse and alleviating the issue of vanishing gradients. Dual Path Networks \cite{chen2017dual} shares common features while maintaining the flexibility to explore new features through dual path architectures. ResNeXt~\cite{xie2017aggregated} introduces a strategy of grouping multiple parallel paths within each residual block, constituting an effective multi-branch structure. Han et al. \cite{han2022multi} design a multi-level back projection structure to extract high-frequency and low-frequency information for image SR reconstruction. While these advancements represent significant progress, there is still considerable scope for further research in the efficient fusion of high- and low-frequency information through multi-branch network architectures.

\section{Architecture}
In this section, we present a detailed explanation of our proposed method, which consists of two subsections: the main architecture and the core Multi-Depth Branch Module (MDBM).

\subsection{Main architecture}

The architectural framework of the proposed MDBN is illustrated in Fig.~\ref{MDBNarc}. The MDBN initially employs a $3\times3$ convolutional layer to the LR image inputs, transforming the RGB channels into a feature space and generating the preliminary shallow features $\mathbf{F}_{\mathrm{initial}}$. 

Subsequently, the method utilizes $6$ blocks of Residual Multi-Depth Branches (RMDBs), which gradually enhance the initial features  $\mathbf{F}_{\mathrm{initial}}$ into more complex deep features $\mathbf{F}_{\mathrm{deep}}$ that are crucial for the reconstruction of HR images. Each RMDB block contains two layers of basic MDBMs, with a residual connection at the ends of each block. In addition, a global residual connection spans the entire RMDB sequence, facilitating the extraction of residual features $\mathbf{F}_{\mathrm{residual}}$.

Following the RMDB processing, the model integrates a trainable upsampler module, which consists of convolutional and PixelShuffle \cite{shi2016PixelShuffle} layers, to upscale the residual features $\mathbf{F}_{\mathrm{residual}}$. This strategic placement of the upsampler minimizes the computational demands. A final $3\times3$ convolutional layer is applied to convert the upsampled features $\mathbf{F}_{\mathrm{upsampled}}$ into the desired HR image outputs.

The formulation of our network can be represented as follows:

\begin{equation}
	\begin{aligned}
		& \mathbf{F}_{\mathrm{initial}}   = \mathcal{C}_{\mathrm{initial}}(\mathbf{I}_{\mathrm{LR}}), \\
		& \mathbf{F}_{\mathrm{deep}}   = \mathcal{M}_{\boldsymbol{\varTheta}}(\mathbf{F}_{\mathrm{initial}}), \\
		& \mathbf{F}_{\mathrm{residual}}   = \mathbf{F}_{\mathrm{initial}} \oplus \mathbf{F}_{\mathrm{deep}}, \\
		& \mathbf{F}_{\mathrm{upsampled}}   = \mathcal{U}_{\boldsymbol{\varPsi}}(\mathbf{F}_{\mathrm{residual}}), \\
		& \mathbf{I}_{\mathrm{SR}}  = \mathcal{C}_{\mathrm{final}}(\mathbf{F}_{\mathrm{upsampled}}), \\
	\end{aligned}	
\end{equation}
where \(\mathbf{I}_{\mathrm{SR}}\) denotes the predicted HR images, and \(\mathbf{I}_{\mathrm{LR}}\) represents the input LR images. The operators \(\mathcal{C}_{\mathrm{initial}}\) and \(\mathcal{C}_{\mathrm{final}}\) stand for the first and last convolutional layers respectively. The stacked RMDBs, parameterized by $\boldsymbol{\varTheta}$, are indicated by the function $\mathcal{M}_{\boldsymbol{\varTheta}}$. While the upsampler module, parameterized by $\boldsymbol{\varPsi}$, is embodied by $\mathcal{U}_{\boldsymbol{\varPsi}}$. Element-wise summation for feature fusion within the network is indicated by the symbol \(\oplus\).

To train these parameters, the Mean Absolute Error (MAE) loss function is employed, defined as:

\begin{equation}
	\mathcal{L}_{\mathrm{MAE}}=\frac{1}{N} \sum_{i=1}^{N}\left\|\mathbf{I}_{\mathrm{SR}}^{(i)}-\mathbf{I}_{\mathrm{HR}}^{(i)}\right\|_{1},
\end{equation}
where \(\mathbf{I}_{\mathrm{HR}}^{(i)}\) represents the ground truth HR image for the \(i\)-th example in the training set, and \(\left\| \cdot \right\|_1\) signifies the  $L1\text{-norm}$.

The model is designed in an end-to-end manner to ensure efficient SISR. In pursuit of optimizing inference speed, the model exclusively employs a stack of $3\times3$ convolutional kernels in conjunction with a single type of activation function.

\begin{table*}[!t]
	\centering
	\caption{Quantitative comparisons of efficient SR networks on commonly used benchmark datasets.}
	\begin{tabular}{|l|c|c|c|c|c|c|}
		\hline
		Methods & Scale & Set5 & Set14 & B100 & Urban100 & Manga109 \\ 
		&  & PSNR / SSIM & PSNR / SSIM & PSNR / SSIM & PSNR / SSIM & PSNR / SSIM \\ \hline
		Bicubic & ~ & 33.66 / 0.9299 & 30.24 / 0.8688 & 29.56 / 0.8431 & 26.88 / 0.8403 & 30.80 / 0.9339 \\ 
		SRCNN \cite{dong2015SRCNN} & ~ & 36.66 / 0.9542 & 32.42 / 0.9063 & 31.36 / 0.8879 & 29.50 / 0.8946 & 35.74 / 0.9661 \\
		FSRCNN \cite{dong2016accelerating} & ~ & 37.00 / 0.9558 & 32.63 / 0.9088 & 31.53 / 0.8920 & 29.88 / 0.9020 & 36.67 / 0.9694 \\ 
		VDSR \cite{kim2016VDSR} & ~ & 37.53 / 0.9587 & 33.03 / 0.9124 & 31.90 / 0.8960 & 30.76 / 0.9140 & 37.22 / 0.9729 \\
		ESPCN \cite{Shi2016ESPCN} & ~ & 36.83 / 0.9564 & 32.40 / 0.9096 & 31.29 / 0.8917 & 29.48 / 0.8975 & - \\		
		LapSRN \cite{lai2017LapSRN} & ~ & 37.52 / 0.9590 & 33.08 / 0.9130 & 31.80 / 0.8950 & 30.41 / 0.9100 & 37.27 / 0.9740 \\ 
		EDSR-baseline \cite{2017EDSRBSFlickr2K} & ~ & 37.99 / 0.9604 & 33.57 / 0.9175 & 32.16 / 0.8994 & 31.98 / 0.9272 & 38.54 / 0.9769 \\ 
		CARN \cite{ahn2018CARN}  & ~ & 37.76 / 0.9590 & 33.52 / 0.9166 & 32.09 / 0.8978 & 31.92 / 0.9256 & - \\ 
		CARN-M \cite{ahn2018CARN} & x2 & 37.53 / 0.9583 & 33.26 / 0.9141 & 31.92 / 0.8960 & 31.23 / 0.9193 & - \\
		IMDN \cite{hui2019IMDN} & ~ & 38.00 / 0.9605     & 33.63 / 0.9177 & \textcolor{blue}{32.19} / 0.8996 & 32.17 / 0.9283 & \textcolor{blue}{38.88} / \textcolor{blue}{0.9774} \\ 
		LAPAR-A \cite{li2020LAPAR} & ~  & 38.01 / 0.9605    & 33.62 / 0.9183 & \textcolor{blue}{32.19} / \textcolor{blue}{0.8999} & 32.10 / 0.9283 & 38.67 / 0.9772 \\ 
		PAN \cite{zhao2020PAN} & ~ & 38.00 / 0.9605     & 33.59 / 0.9181 & 32.18 / 0.8997 & 32.01 / 0.9273 & 38.70 / 0.9773 \\ 
		A$^{2}$F-M \cite{wang2020accv} & ~ & \textcolor{blue}{38.04} / \textcolor{blue}{0.9607}     & \textcolor{blue}{33.67} / \textcolor{blue}{0.9184} & 32.18 / 0.8996 & \textcolor{blue}{32.27} / \textcolor{blue}{0.9294} & 38.87 / \textcolor{blue}{0.9774} \\ 
		SMSR \cite{wang2021SMSR} & ~ & 38.00 / 0.9601    & 33.64 / 0.9179 & 32.17 / 0.8990 & 32.19 / 0.9284 & 38.76 / 0.9771 \\ 
		ShuffleMixer \cite{sun2022ShuffleMixerNIPS} & ~ & 38.01 / 0.9606     & 33.63 / 0.9180 & 32.17 / 0.8995 & 31.89 / 0.9257 & 38.83 / \textcolor{blue}{0.9774} \\ 
		SAFMN \cite{sun2023SAFM} & ~ & 38.00 / 0.9605     & 33.54 / 0.9177 & 32.16 / 0.8995 & 31.84 / 0.9256 & 38.71 / 0.9771 \\ 
		\textbf{MDBN (Ours)} & ~ & \textbf{38.13} / \textbf{0.9614} & \textbf{33.89} / \textbf{0.9207} & \textbf{32.28} / \textbf{0.9017} & \textbf{32.54} / \textbf{0.9322} & \textbf{39.05} / \textbf{0.9782} \\ \hline 
		Bicubic & ~ & 30.39 / 0.8682 & 27.55 / 0.7742 & 27.21 / 0.7385 & 24.46 / 0.7349 & 26.95 / 0.8556 \\ 
		SRCNN \cite{dong2015SRCNN} & ~ & 32.75 / 0.9090  & 29.28 / 0.8209 &  28.41 / 0.7863 & 26.24 / 0.7989 & 30.59 / 0.9107 \\
		FSRCNN \cite{dong2016accelerating} & ~ & 33.16 / 0.9140 & 29.43 / 0.8242 & 28.53 / 0.7910 & 26.43 / 0.8080 & 30.98 / 0.9212 \\ 
		VDSR \cite{kim2016VDSR} & ~ & 33.66 / 0.9213 & 29.77 / 0.8314 & 28.82 / 0.7976 & 27.14 / 0.8279 & 32.01 / 0.9310 \\
		EDSR-baseline \cite{2017EDSRBSFlickr2K} & ~ & 34.37 / 0.9270 & 30.28 / 0.8417 & 29.09 / 0.8052 & 28.15 / 0.8527 & 33.45 / 0.9439 \\ 
		CARN \cite{ahn2018CARN} & ~ & 34.29 / 0.9255 & 30.29 / 0.8407 & 29.06 / 0.8034 & 28.06 / 0.8493 & - \\ 
		CARN-M \cite{ahn2018CARN} & ~ & 33.99 / 0.9236 & 30.08 / 0.8367 & 28.91 / 0.8000 & 27.55 / 0.8385 & - \\
		IMDN \cite{hui2019IMDN} & x3 & 34.36 / 0.9270 & 30.32 / 0.8417 & 29.09 / 0.8046 & 28.17 / 0.8519 & 33.61 / 0.9445 \\ 
		LAPAR-A \cite{li2020LAPAR} & ~ & 34.36 / 0.9267 & 30.34 / 0.8421 & 29.11 / \textcolor{blue}{0.8054} & 28.15 / 0.8523 & 33.51 / 0.9441 \\ 
		PAN \cite{zhao2020PAN} & ~ & 34.40 / 0.9271 & 30.36 / 0.8423 & 29.11 / 0.8050 & 28.11 / 0.8511 & 33.61 / 0.9448 \\
		A$^{2}$F-M \cite{wang2020accv} & ~ & \textcolor{blue}{34.50} / \textcolor{blue}{0.9278}     & \textcolor{blue}{30.39} / \textcolor{blue}{0.8427} & 29.11 / \textcolor{blue}{0.8054} & \textcolor{blue}{28.28} / \textcolor{blue}{0.8546} & 33.66 / \textcolor{blue}{0.9453} \\ 
		SMSR \cite{wang2021SMSR} & ~ & 34.40 / 0.9270 & 30.33 / 0.8412 & 29.10 / 0.8050 & 28.25 / 0.8536 & 33.68 / 0.9445 \\ 
		ShuffleMixer \cite{sun2022ShuffleMixerNIPS} & ~ & 34.40 / 0.9272 & 30.37 / 0.8423 & \textcolor{blue}{29.12} / 0.8051 & 28.08 / 0.8498 & \textcolor{blue}{33.69} / 0.9448 \\ 
		SAFMN \cite{sun2023SAFM} & ~ & 34.34 / 0.9267 & 30.33 / 0.8418 & 29.08 / 0.8048 & 27.95 / 0.8474 & 33.52 / 0.9437 \\ 
		\textbf{MDBN (Ours)} & ~ & \textbf{34.53} / \textbf{0.9284} & \textbf{30.46} / \textbf{0.8448} & \textbf{29.20} / \textbf{0.8088} & \textbf{28.47} / \textbf{0.8586} & \textbf{33.90} / \textbf{0.9468} \\ \hline
		Bicubic & ~ & 28.42 / 0.8104 & 26.00 / 0.7027 & 25.96 / 0.6675 & 23.14 / 0.6577 & 24.89 / 0.7866 \\ 
		SRCNN \cite{dong2015SRCNN} & ~ & 30.48 / 0.8628 & 27.49 / 0.7503 & 26.90 / 0.7101 &  24.52 / 0.7221 &  27.66 / 0.8505 \\
		FSRCNN \cite{dong2016accelerating} & ~ & 30.71 / 0.8657 & 27.59 / 0.7535 & 26.98 / 0.7150 &  24.62 / 0.7280 & 27.90 / 0.8517 \\ 
		VDSR \cite{kim2016VDSR} & ~ & 31.35 / 0.8838 & 28.01 / 0.7674 & 27.29 / 0.7251 & 25.18 / 0.7524 & 28.83 / 0.8809 \\
		ESPCN \cite{Shi2016ESPCN} & ~ & 30.52 / 0.8697 & 27.42 / 0.7606 & 26.87 / 0.7216 &  24.39 / 0.7241 & - \\
		LapSRN \cite{lai2017LapSRN} & ~ & 31.54 / 0.8850 & 28.19 / 0.7720 & 27.32 / 0.7280 & 25.21 / 0.7560 & 29.09 / 0.8845 \\ 
		EDSR-baseline \cite{2017EDSRBSFlickr2K} & ~ & 32.09 / 0.8938 & 28.58 / 0.7813 & 27.57 / 0.7357 & 26.04 / 0.7849 & 30.35 / 0.9067 \\ 
		CARN \cite{ahn2018CARN} & ~ & 32.13 / 0.8937 & 28.60 / 0.7806 & 27.58 / 0.7349 & 26.07 / 0.7837 & - \\ 
		CARN-M \cite{ahn2018CARN} & x4 & 31.92 / 0.8903 & 28.42 / 0.7762 & 27.44 / 0.7304 &  25.62 / 0.7694 & - \\
		IMDN \cite{hui2019IMDN} & ~ & 32.21 / 0.8948 & 28.58 / 0.7811 & 27.56 / 0.7353 & 26.04 / 0.7838 & 30.45 / 0.9075 \\ 
		LAPAR-A \cite{li2020LAPAR} & ~ & 32.15 / 0.8944 & 28.61 / 0.7818 & \textcolor{blue}{27.61} / \textcolor{blue}{0.7366} & 26.14 /  0.7871 & 30.42 / 0.9074 \\ 
		PAN \cite{zhao2020PAN} & ~ & 32.13 / 0.8948 & 28.61 / 0.7822 & 27.59 / 0.7363 & 26.11 / 0.7854 & 30.51 / 0.9095 \\ 
		A$^{2}$F-M \cite{wang2020accv} & ~ & \textcolor{blue}{32.28} / \textcolor{blue}{0.8955}     & 28.62 / \textcolor{blue}{0.7828} & 27.58 / 0.7364 & \textcolor{blue}{26.17} / \textcolor{blue}{0.7892} & 30.57 / \textcolor{blue}{0.9100} \\ 
		SMSR \cite{wang2021SMSR} & ~ & 32.12 / 0.8932 & 28.55 / 0.7808 & 27.55 / 0.7351 & 26.11 / 0.7868 & 30.54 / 0.9085 \\ 
		ShuffleMixer \cite{sun2022ShuffleMixerNIPS} & ~ & 32.21 / 0.8953 & \textcolor{blue}{28.66} / 0.7827 & \textcolor{blue}{27.61} / \textcolor{blue}{0.7366} & 26.08 / 0.7835 & \textcolor{blue}{30.65} / 0.9093 \\ 
		SAFMN \cite{sun2023SAFM} & ~ & 32.18 / 0.8948 & 28.60 / 0.7813 & 27.58 / 0.7359 & 25.97 / 0.7809 & 30.43 / 0.9063 \\ 
		\textbf{MDBN (Ours)} & ~ & \textbf{32.31} / \textbf{0.8973} & \textbf{28.72} / \textbf{0.7853} & \textbf{27.67} / \textbf{0.7408} & \textbf{26.34} / \textbf{0.7938} & \textbf{30.80} / \textbf{0.9123} \\ \hline
	\end{tabular}
	
	\label{table1comp}
\end{table*}

\subsection{Multi-depth branch module}

The architecture and feature map visualizations of the MDBM are depicted in Fig.~\ref{MDBNFM}(a). The high-frequency branch of the MDBM consists of two sequential $3\times3$ convolutional layers, separated by a Gaussian Error Linear Unit (GELU) \cite{2023GELUs} activation function. In contrast, the low-frequency branch is composed of a single $3\times3$ convolutional layer. The outputs of both branches are fused together by element-wise addition, followed by a GELU activation function before being fed into the subsequent MDBM.

The processing of the input feature $\mathbf{F}_{\mathrm{input}}$ in the MDBM is mathematically represented as:

\begin{equation}
	\begin{split}
		& \mathbf{F}_{\mathrm{hf}}   =  \mathcal{C}_{\mathrm{up_{2}}}(\mathcal{G}(\mathcal{C}_{\mathrm{up_{1}}}(\mathbf{F}_{\mathrm{input}}))), \\
		& \mathbf{F}_{\mathrm{lf}}   = \mathcal{C}_{\mathrm{lf}}(\mathbf{F}_{\mathrm{input}}), \\
		& \mathbf{F}_{\mathrm{output}}   = \mathcal{G}(\mathbf{F}_{\mathrm{hf}} \oplus \mathbf{F}_{\mathrm{lf}}). \\
	\end{split}	
\end{equation}

Here, $\mathcal{C}{(\cdot)}$ denotes a $3\times3$ convolutional operator and $\mathcal{G}$ represents the GELU activation function. The feature maps $\mathbf{F}_{\mathrm{hf}}$ and $\mathbf{F}_{\mathrm{lf}}$ are derived from the high-frequency and low-frequency branches respectively.

From the visualized feature maps in Fig.~\ref{MDBNFM}(a), one can see that $\mathbf{F}_{\mathrm{hf}}$ captures detailed textures essential for image SR, while $\mathbf{F}_{\mathrm{lf}}$ focuses on outlining contours. By integrating $\mathbf{F}_{\mathrm{hf}}$ and $\mathbf{F}_{\mathrm{lf}}$ to form the comprehensive feature $\mathbf{F}_{\mathrm{output}}$, our model achieves the efficient generation of high-quality HR images. This integration not only preserves high-frequency details but also ensures semantic coherence, contributing to the overall quality of the resulting images.

Figure~\ref{MDBNFM}(b) illustrates the normalized Fourier spectra of feature maps from different branches. The computation of these normalized spectra is defined as:

\begin{table*}[!t]
	\centering
	\caption{Comparisons of inference time and performance on B100 dataset on $\times 4$ SR. Inference time is the average test time on an NVIDIA RTX 3090 GPU over a set of 50 LR images with dimensions of 320 × 180 pixels.}
	\begin{tabular}{|l|l|c|c|}
		\hline
		Category & Methods  & Inference Time (ms) & B100 (PSNR / SSIM) \\ \hline
		CNN-based & EDSR-baseline \cite{2017EDSRBSFlickr2K}  & 19.81 & 27.57 / 0.7357 \\
		~ & CARN \cite{ahn2018CARN}  & 18.90 & 27.58 / 0.7349 \\ 
		~ & IMDN \cite{hui2019IMDN}  & 10.22 & 27.56 / 0.7353 \\ 
		~ & LAPAR-A \cite{li2020LAPAR}  & 24.91 & 27.61 / 0.7366 \\
		~ & SMSR \cite{wang2021SMSR}  & 22.48 & 27.55 / 0.7351 \\
		~ & ShuffleMixer \cite{sun2022ShuffleMixerNIPS}  & 20.16 & 27.61 / 0.7366 \\
		~ & SAFMN \cite{sun2023SAFM}  & 10.71 & 27.58 / 0.7359\\ 
		~ & MDBN (Ours)  & 10.69 & \textcolor{blue}{27.67} / \textbf{0.7408} \\ \hline
		Transformer-based & SwinIR-light \cite{liang2021swinir}  & 180.36 & \textbf{27.69} / \textcolor{blue}{0.7406} \\
		~ & SCET \cite{zou2022selftransformer} & 12.84 & \textcolor{blue}{27.67} / 0.7390 \\
		~ & ELAN-light \cite{zhang2022ELAN}  & 41.25 & \textbf{27.69} / \textcolor{blue}{0.7406} \\
		~ & HPINet-S \cite{liu2023coarse} & 142.57 & \textbf{27.69} / - \\ \hline
		
	\end{tabular}
	
	\label{table2time}
\end{table*}

\begin{equation}\label{eq::norm}
\begin{split}
&\mathcal{F}_{\min}  = \min \Big( \big|\mathcal{FFT}(\mathbf{F}_{\mathrm{hf}}) \big|,  \big|\mathcal{FFT}(\mathbf{F}_{\mathrm{lf}}) \big| \Big), \\
&\mathcal{F}_{\max}  = \max \Big( \big|\mathcal{FFT}(\mathbf{F}_{\mathrm{hf}}) \big|,  \big|\mathcal{FFT}(\mathbf{F}_{\mathrm{lf}}) \big| \Big), \\
&\hat{\mathbf{F}}_{\mathrm{hf}}=\frac{ \big|\mathcal{FFT}(\mathbf{F}_{\mathrm{hf}}) \big|-\mathcal{F}_{\min}}{\mathcal{F}_{\max}-\mathcal{F}_{\min}}, \\
&\hat{\mathbf{F}}_{\mathrm{lf}}=\frac{ \big|\mathcal{FFT}(\mathbf{F}_{\mathrm{lf}}) \big|-\mathcal{F}_{\min}}{\mathcal{F}_{\max}-\mathcal{F}_{\min}},
\end{split} 
\end{equation}
where $\mathcal{FFT}(\cdot)$ represents the Fast Fourier Transform (FFT) operator, and $|\cdot|$ denotes the magnitude of a complex number.

In Fig.~\ref{MDBNFM}(b), regions with more larger values in the center show more low-frequency content. From $\hat{\mathbf{F}}_{\mathrm{hf}}$ it can be further proven that the deeper branch mainly predicts the high-frequency information and captures fine-grained object details. Meanwhile, $\hat{\mathbf{F}}_{\mathrm{lf}}$ reveals that the shallower branch contains more low-frequency semantic understanding and delineates broader object contours.   

\begin{figure*}[!t]
	\centering
	\includegraphics[width=0.98\textwidth]{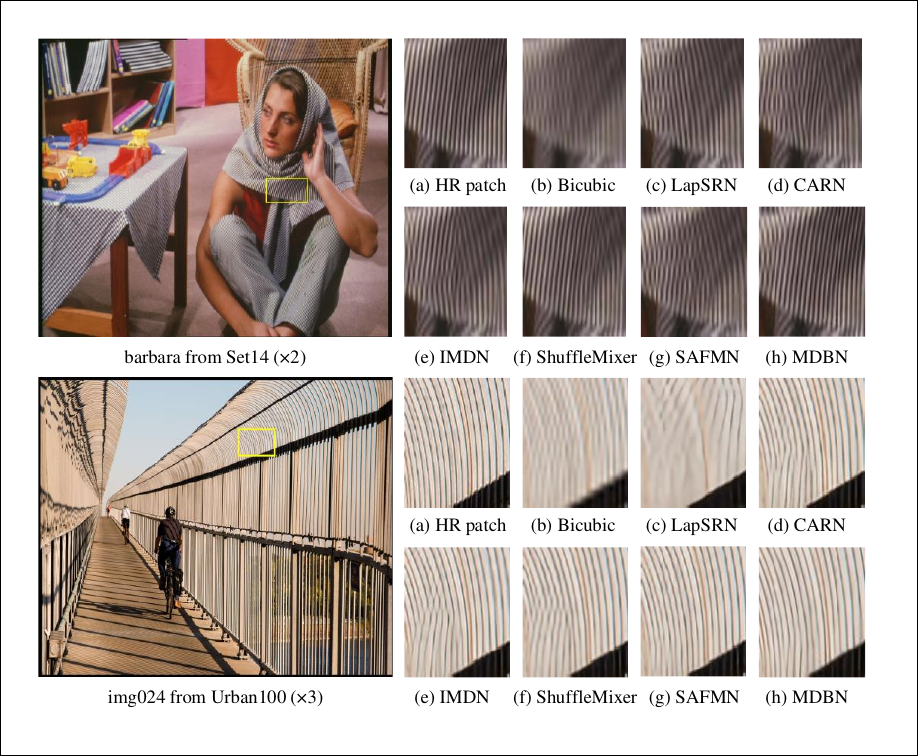} 
	\caption{Visual comparisons for $\times2$ and $\times3$ SR. (a) HR patch of ground truth images, (b) Bicubic, (c) LapSRN \cite{lai2017LapSRN}, (d) CARN \cite{ahn2018CARN}, (e) IMDN \cite{hui2019IMDN}, (f) ShuffleMixer \cite{sun2022ShuffleMixerNIPS}, (g) SAFM \cite{sun2023SAFM}, (h) MDBN (ours).}
	\label{MDBNX23}
\end{figure*}

\begin{figure*}[!t]
	\centering
	\includegraphics[width=0.98\textwidth]{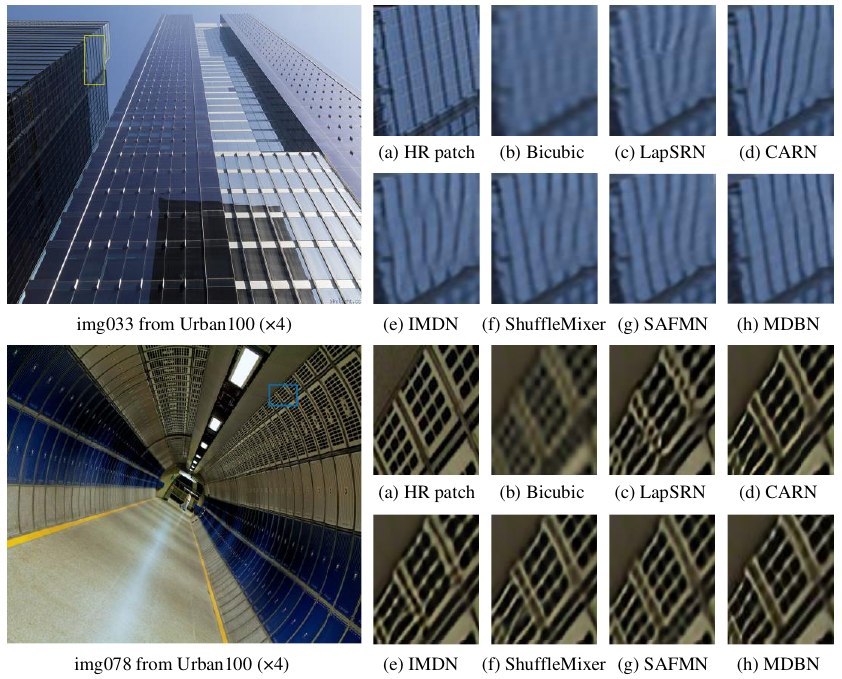} 
	\caption{Visual comparisons for $\times4$ SR. (a) HR patch of ground truth images, (b) Bicubic, (c) LapSRN \cite{lai2017LapSRN}, (d) CARN \cite{ahn2018CARN}, (e) IMDN \cite{hui2019IMDN}, (f) ShuffleMixer \cite{sun2022ShuffleMixerNIPS}, (g) SAFM \cite{sun2023SAFM}, (h) MDBN (ours).}
	\label{MDBNX4}
\end{figure*}

\section{Experiments}

In this section, quantitative and qualitative analyses are conducted to validate the effectiveness of the proposed MDBN method.

\subsection{Experimental setting}
\paragraph{Datasets.} Consistent with previous studies \cite{li2020LAPAR, sun2022ShuffleMixerNIPS, sun2023SAFM}, we use the DIV2K \cite{2017DIV2K} and Flickr2K \cite{2017EDSRBSFlickr2K} datasets for training. The LR images are generated through bicubic downscaling of the corresponding HR images. For testing, we utilize five widely recognized benchmark datasets:  Set5 \cite{2012Set5}, Set14 \cite{2012Set14}, Urban100 \cite{2015UR100}, B100 \cite{2010B100}, and Manga109 \cite{2017manga109}. 

\paragraph{Implementation Details.} Our model is implemented using the BasicSR framework \cite{basicsr} within the PyTorch v1.12 environment, and is executed on an NVIDIA RTX 3080 GPU. A batch size of 16 is used for training. Data augmentation techniques, including random horizontal flips and rotations, are applied to the input patches. The Adam optimizer \cite{kingma2014adam} is used with the settings $\beta_{1} = 0.9$ and $\beta_2 = 0.99$. The number of iterations is set to 1,800,000. The learning rate is initially set at 2e-4, with a minimum threshold of 1e-7, and is updated following the Cosine Annealing scheme \cite{loshchilov2016cosine}. All our results presented in this paper are obtained directly from the trained networks, without any additional post-processing steps. Our code is available at \url{https://github.com/thy960112/MDBN}.

\subsection{Quantitative assessment}

The quality of the reconstructed images is evaluated using the Peak Signal to Noise Ratio (PSNR) and the Structural Similarity Index (SSIM). These metrics are computed on the Y channel of the images converted to the YCbCr color space. Higher PSNR represents less pixel-level discrepancies and larger SSIM  means better human visual perception. For a more detailed description of PSNR and SSIM, readers can refer to  \cite{moser2023hitchhiker}.

Table~\ref{table1comp} shows the quantitative comparison of our proposed MDBN and the SOTA efficient SR methods: SRCNN \cite{dong2015SRCNN}, FSRCNN \cite{dong2016accelerating},  VDSR \cite{kim2016VDSR}, ESPCN \cite{Shi2016ESPCN}, LapSRN \cite{lai2017LapSRN}, EDSR baseline \cite{2017EDSRBSFlickr2K}, CARN \cite{ahn2018CARN}, IMDN \cite{hui2019IMDN}, LAPAR \cite{li2020LAPAR}, PAN \cite{zhao2020PAN}, A$^{2}$F-M \cite{wang2020accv}, SMSR \cite{wang2021SMSR}, ShuffleMixer \cite{sun2022ShuffleMixerNIPS}, and SAFM \cite{sun2023SAFM}. The results demonstrate that our method surpasses competing approaches across all benchmark datasets for $\times 2$, $\times 3$, and $\times 4$ super-resolution (SR) tasks.

Specifically, the MDBN outperforms the second-best methods across all three scales, with a margin of $0.17\mathrm{dB}$ to $0.27\mathrm{dB}$ on the Urban100 and $0.15\mathrm{dB}$ to $0.21\mathrm{dB}$ on Manga109. These two datasets are characterized by their rich details and structural semantics. This performance underscores the effectiveness of our asymmetric architecture in accurately predicting high-frequency details, guided by contextual understanding.

In the scenario of efficient SR networks, inference time is of critical importance. Table~\ref{table2time} illustrates a comparative analysis of inference time and performance on B100 dataset for $\times 4$ SR. Following the previous work \cite{sun2023SAFM}, inference time is averaged over a set of 50 LR images with dimensions of $320 \times 180$ pixels. These tests are conducted using an NVIDIA RTX 3090 GPU.

\begin{table*}[!t]
	\centering
	\caption{Ablation study for $\times2$ SR for our proposed Multi-Depth Branch Network (MDBN) on Set5 \cite{2012Set5} and Set14 \cite{2012Set14} datasets.}
	\begin{tabular}{|l|c|c|c|}
		\hline
		Ablation & Variants & Set5 & Set14 \\ 
		~ & ~ & PSNR / SSIM & PSNR / SSIM \\ \hline
		Baseline & - & 38.13 / 0.9614 & 33.88 / 0.9207 \\ \hline
		Main module & 2 MDBM $\times$ 6 RMDB $\to$ 1 MDBM $\times$ 12 RMDB  & 38.07 / 0.9600 & 33.83 / 0.9175 \\ \hline
		MDBM & 2 $\times$ Conv 3$\times$3 $\to$ 1 $\times$ Conv 5$\times$5 & 37.94 / 0.9581 & 33.62 / 0.9161 \\
		~ & GELU $\to$ LeakyReLU & 37.96 / 0.9609 & 33.63 / 0.9172 \\ \hline
		Learning rate & Cosine Annealing $\to$ MultiStep & 38.09 / 0.9613 & 33.81 / 0.9199 \\ 
		~ & lr 2e-4 $\to$ lr 1e-3 & 37.09 / 0.9503 & 33.10 / 0.9051 \\ \hline
		Normalization & w/o Norm $\to$ BatchNorm & 17.55 / 0.2906 & 16.78 / 0.2297 \\ 
		~ & w/o Norm $\to$  Input/Output Norm & 37.95 / 0.9608 & 33.60 / 0.9176 \\ \hline
		Datasets & DIV2K and Flickr2K $\to$  DIV2K & 38.08 / 0.9612 & 33.82 / 0.9191 \\ \hline

	\end{tabular}
	
	\label{table3analysis}
\end{table*}

From Table~\ref{table1comp} and \ref{table2time}, we can see that our MDBN achieves superior performance over the EDSR-baseline across all benchmark datasets while demonstrating approximately twice the inference speed. This indicates that the branch architecture in our model can produce better results in a significantly shorter inference time.

Analysis of the data in Table~\ref{table2time} reveals that most transformer-based methods marginally exceed our MDBN by $0.02\mathrm{dB}$ on the B100 dataset. However, the SSIM achieved by MDBN slightly surpasses that of transformer-based methods. This indicates the superior ability of MDBN to preserve structural coherence in images. The inference time of most transformer-based methods is significantly longer, ranging from $1.2\times$ to $18\times$ slower than that of MDBN. Notably, SCET \cite{zou2022selftransformer} only uses one transformer layer and the remaining layers are based on CNN. This highlights the advantage of MDBN in terms of computational efficiency, making it suitable for low-power devices.

\begin{figure*}[!t]
	\centering
	\includegraphics[width=0.9\textwidth]{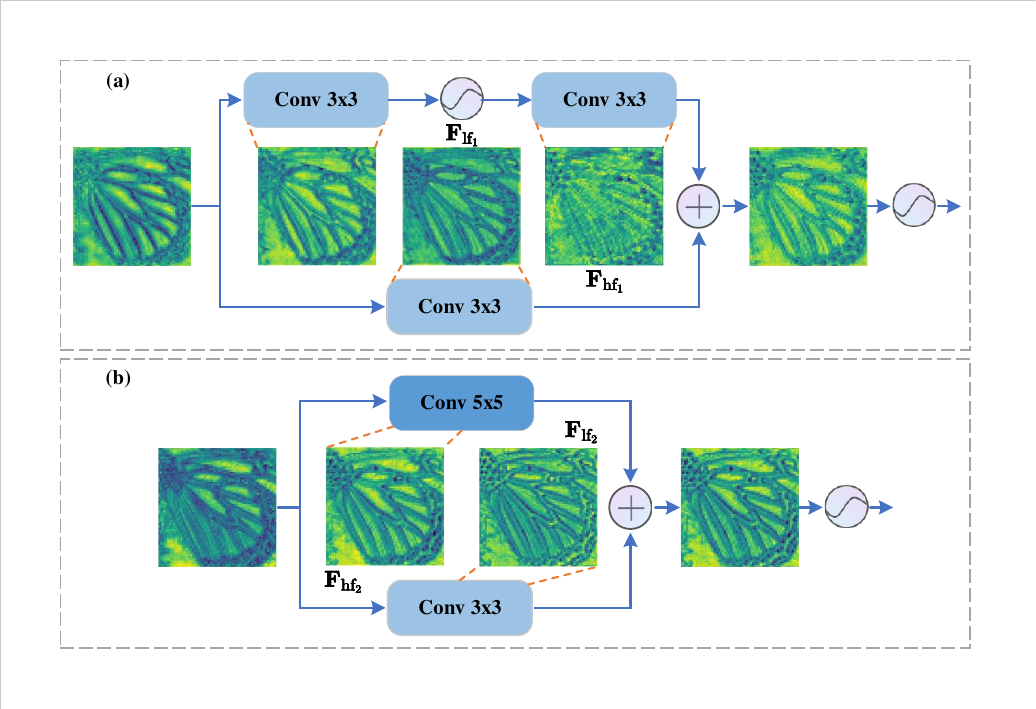} 
	\caption{Comparison ($\times 4$ SR) of feature maps of (a) Multi-Depth Branch Module (MDBM) and its counterpart (b), wherein the two stacked $3\times 3$ convolutional layers are replaced by a single $5\times 5$ kernel.}
	\label{MDBNCPFM}
\end{figure*}

\begin{figure*}[!t]
	\centering
	\includegraphics[width=0.9\textwidth]{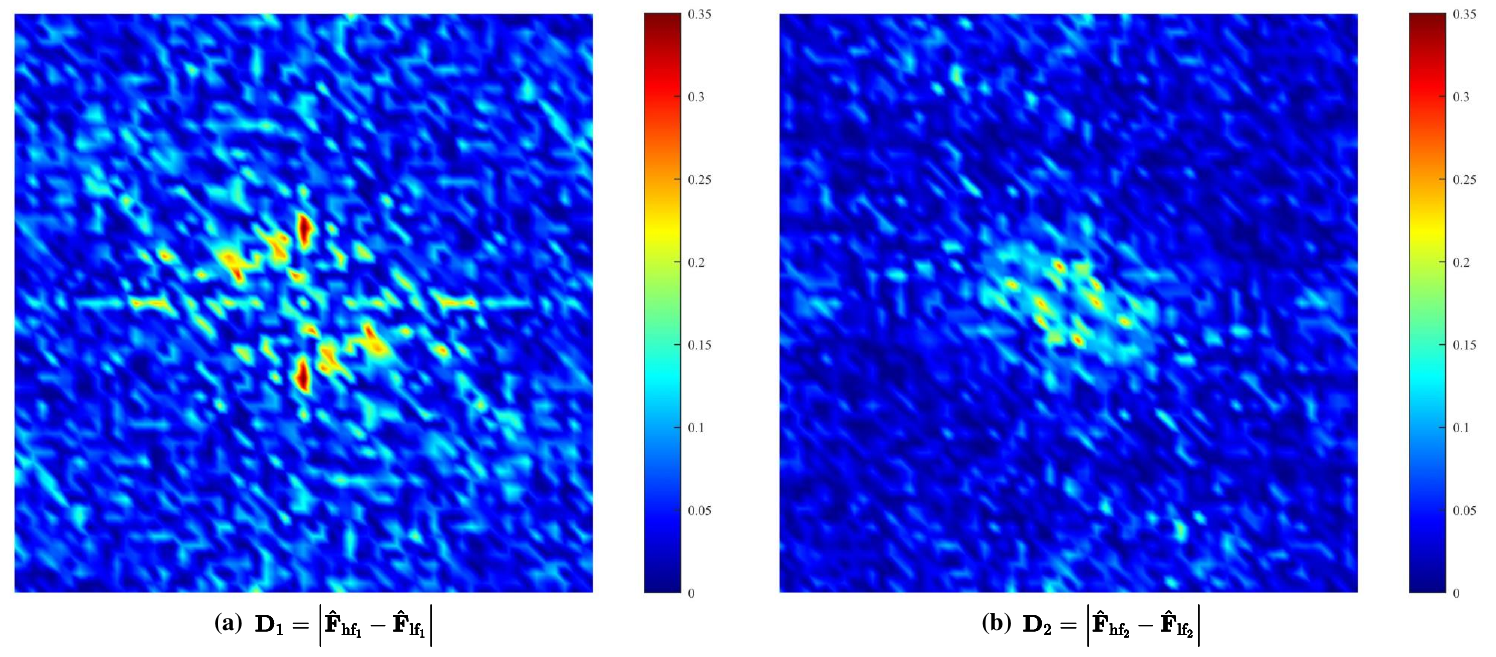} 
	\caption{Visualized comparison ($\times 4$ SR) between (a) our proposed MDBM and (b) branches where the two stacked $3\times 3$ convolutional layers are replaced by a single $5\times 5$ kernel. The values in the figures represent the absolute difference in amplitude of the normalized Fourier spectra for the feature maps of different branches. The difference of our proposed MDBM is larger than its counterpart, which means less redundancy in feature extraction.}
	\label{MDBNCPFFT}
\end{figure*}

\subsection{Qualitative assessment}

In addition to the quantitative assessments, we also present qualitative comparisons of our MDBN against SOTA efficient SR methods, including LapSRN \cite{lai2017LapSRN}, CARN \cite{ahn2018CARN}, IMDN \cite{hui2019IMDN}, ShuffleMixer \cite{sun2022ShuffleMixerNIPS}, and SAFM \cite{sun2023SAFM}. 
Figure~\ref{MDBNX23} presents visual comparisons on the Set14 and Urban100 datasets for $\times2$ and $\times3$ SR scales. Figure~\ref{MDBNX4} displays comparison results on the Urban100 dataset for $\times4$ SR. 

The qualitative comparison results illustrate that the MDBN is adept at producing images that are both semantically coherent and visually realistic. This emphasizes the effectiveness of the shallower branch in our architecture, which provides essential structural guidance to the deeper branch, thereby enriching fine-grained details in SR tasks. Such results underscore the overall efficacy of the MDBN model in handling complex image enhancement challenges.

\section{Analysis and discussion}

\subsection{Ablation study}

To determine the effectiveness of each component within our network, we conduct an ablation analysis for $\times2$ SR. The results of this analysis are presented in Table~\ref{table3analysis}. A key modification in the main architecture is to change the residual connection in each RMDB block, reducing the number of MDBMs from $2$ to $1$, while maintaining the same total number of layers. This change lead to a marginal decrease in both PSNR and SSIM values compared to the baseline model. Additionally, replacing the two $3\times 3$ convolutional layers in the high-frequency branch of the MDBM with a $5\times 5$ convolutional layer results in an increase in parameters and a decrease in performance. These findings further highlight the effectiveness and efficiency of our original MDBM architecture.

Changes in activation functions, learning rate control strategies, the use of input/output normalization and datasets have minimal influence on the final results. However, different initial values of the learning rate are observed to significantly affect the outputs. In particular, batch normalization, a technique commonly used in other computer vision tasks, is found to be incompatible with the training of SR networks. 

\subsection{Feature map visualizations}

For a better understanding of the proposed network, Fig~\ref{MDBNCPFM}(a) shows the feature maps of the third RMDB in the proposed network on $\times4$ SR. Meanwhile, Fig~\ref{MDBNCPFM}(b) presents the feature maps of its counterpart, wherein the two stacked $3\times 3$ convolutional layers are replaced by a single $5\times 5$ convolutional layer. 

It is shown in Fig~\ref{MDBNCPFM}(a) that the high-frequency branch is adept at restoring intricate image details and the low-frequency branch primarily focuses on rectifying contours. However, in Fig~\ref{MDBNCPFM}(b), the disparity between feature maps of different branches is negligible. 

\subsection{Fourier spectral analysis of feature maps}

In this section, we propose a Fourier spectral analysis method to further validate the efficacy of our proposed MDBN. This technique is readily adaptable for analyzing various network architectures. 

Our approach involves comparing the absolute differences in magnitude of the normalized Fourier spectra for the feature maps of different branches from Fig.~\ref{MDBNCPFM}. We first perform FFT on these features and obtain the maximum and minimum magnitudes of the spectral maps:

\begin{equation}
	\begin{split}
		&\mathcal{F}_{\mathrm{hf_{1}}} = \mathcal{FFT}(\mathbf{F}_{\mathrm{hf_{1}}}), \\
		&\mathcal{F}_{\mathrm{hf_{2}}} = \mathcal{FFT}(\mathbf{F}_{\mathrm{hf_{2}}}),  \\
		&\mathcal{F}_{\mathrm{lf_{1}}} = \mathcal{FFT}(\mathbf{F}_{\mathrm{lf_{1}}}), \\
		&\mathcal{F}_{\mathrm{lf_{2}}} = \mathcal{FFT}(\mathbf{F}_{\mathrm{lf_{2}}}),\\
		&\mathcal{F}_{\min} = \min \Big( \big|\mathcal{F}_{\mathrm{hf_{1}}} \big|, \big|\mathcal{F}_{\mathrm{hf_{2}}} \big|, \big|\mathcal{F}_{\mathrm{lf_{1}}} \big|, \big|\mathcal{F}_{\mathrm{lf_{2}}} \big| \Big), \\  
		&\mathcal{F}_{\max} = \max \Big( \big|\mathcal{F}_{\mathrm{hf_{1}}} \big|, \big|\mathcal{F}_{\mathrm{hf_{2}}} \big|, \big|\mathcal{F}_{\mathrm{lf_{1}}} \big|, \big|\mathcal{F}_{\mathrm{lf_{2}}} \big| \Big). \\
	\end{split}
\end{equation}

Subsequently, the spectral maps are collectively normalized as follows:

\begin{equation}
	\begin{split}  
		\hat{\mathbf{F}}_{\mathrm{hf_{1}}}&=\frac{ \big|\mathcal{F}_{\mathrm{hf_{1}}} \big|-\mathcal{F}_{\min}}{\mathcal{F}_{\max}-\mathcal{F}_{\min}}, \quad  
		\hat{\mathbf{F}}_{\mathrm{lf_{1}}}&=\frac{ \big|\mathcal{F}_{\mathrm{lf_{1}}} \big|-\mathcal{F}_{\min}}{\mathcal{F}_{\max}-\mathcal{F}_{\min}}, \\
		\hat{\mathbf{F}}_{\mathrm{hf_{2}}}&=\frac{ \big|\mathcal{F}_{\mathrm{hf_{2}}} \big|-\mathcal{F}_{\min}}{\mathcal{F}_{\max}-\mathcal{F}_{\min}}, \quad  
		\hat{\mathbf{F}}_{\mathrm{lf_{2}}}&=\frac{ \big|\mathcal{F}_{\mathrm{lf_{2}}} \big|-\mathcal{F}_{\min}}{\mathcal{F}_{\max}-\mathcal{F}_{\min}}. \\
	\end{split}  
\end{equation}

Finally, the absolute difference in magnitude of the normalized Fourier spectra can be calculated by:

\begin{equation}  
	\begin{split}
		\mathbf{D}_{1} &= \big | \hat{\mathbf{F}}_{\mathrm{hf_{1}}} - \hat{\mathbf{F}}_{\mathrm{lf_{1}}} \big |, \\
		\mathbf{D}_{2} &= \big | \hat{\mathbf{F}}_{\mathrm{hf_{2}}} - \hat{\mathbf{F}}_{\mathrm{lf_{2}}} \big |,
	\end{split}
\end{equation}

where $\mathbf{D}_{1}$ and $\mathbf{D}_{2}$ represent spectral differences of feature maps from Fig.~\ref{MDBNCPFM}(a) and Fig.~\ref{MDBNCPFM}(b) respectively. As shown in Fig.~\ref{MDBNCPFFT}, the larger values in $\mathbf{D}_{1}$ mean that the different branches of MDBM are able to process different frequency information separately compared to other branch networks, and the difference is obvious. The distribution of brighter values in $\mathbf{D}_{1}$ is more expansive, indicating the differences are not confined to the low-frequency central region but also extend to the surrounding high-frequency areas. This extensive variance suggests that the MDBN can capture a wide range of frequency information, efficiently covering both high-frequency details and low-frequency contour information. Furthermore, this attribute demonstrates that the MDBM can effectively reduce the information redundancy and improve the computational efficiency of the model during the feature extraction process.

The Fourier spectral analysis focuses on the frequency domain, revealing how the network handles different frequency components. This analysis technique can also quantify the spectral differences between the feature maps of different layers or branches of the network. By combining visualization and numerical analysis, this spectral technique offers deeper insights into how neural networks process and represent information in the frequency domain. Such insights can guide the design of more efficient and effective neural network architectures.

\section{Conclusion}

In this paper, we present an asymmetric MDBN model that focuses on efficiently enhancing high-frequency details in LR images while maintaining semantic coherence. The core MDBM architecture features branches of different depths, designed to capture high- and low-frequency information simultaneously and efficiently. Its hierarchical structure allows the deeper branch to accumulate fine-grained local details, contextually guided by the shallower branch. The effectiveness and rationality of this design are illustrated through feature map visualizations and proposed novel Fourier spectral analysis methods. The model shows a more pronounced spectral differentiation between branches compared to existing branch networks, indicating a reduction in feature redundancy and a more effective integration of high- and low-frequency information. Extensive qualitative and quantitative evaluations on various datasets confirm the ability of the proposed model in generating structurally consistent and visually realistic HR images, achieving SOTA results with a very fast inference speed.

\section*{Declaration of Competing Interest}
The authors have no competing interests to declare that are relevant to the content of this article.

\section*{Data availability}
To ensure that the results are reproducible, the source code is publicly available at \url{https://github.com/thy960112/MDBN}.

\section*{Acknowledgements}
The authors gratefully acknowledge the financial support of the Natural Science Foundation of China (Grant 61925603).

\bibliographystyle{cas-model2-names}

\bibliography{cas-refs}

\end{document}